\documentclass{aastex701}
\usepackage{grffile}          
\graphicspath{{figs/}}        

\usepackage{graphicx}
\usepackage{amsmath,amssymb}
\usepackage{booktabs}
\usepackage{txfonts}
\usepackage{gensymb}
\usepackage{multirow}
\usepackage{longtable}
\usepackage{comment}
\newlength{\panelH}
\usepackage{xcolor}
\definecolor{modelbg}{RGB}{63,63,63}

\begin{document}

\title{Physical Characteristics of the Asteroid (469219) Kamo'oalewa as a target of the Chinese Tianwen-2 mission}


\author[0000-0002-7421-0532,gname=Xiaobin,sname=Wang]{Xiaobin Wang}
\affiliation{Yunnan Observatories, Chinese Academy of Sciences, 
Kunming, 650216, China}
\affiliation{Key Laboratory for the Structure and Evolution of Celestial Objects, Chinese Academy of Sciences, Kunming 650216, China}
\affiliation{University of Chinese Academy of Sciences, Beijing 100049, China}
\email[show]{wangxb@ynao.ac.cn}

\author{Karri Muinonen}
\affiliation{Department of Physics, University of Helsinki, P.O. box 64, FI-00014 U. Helsinki, Finland}
\email{Karri.muinonen@helsinki.fi}

\author{Shenghong Gu}
\affiliation{Yunnan Observatories, Chinese Academy of Sciences, 
Kunming, 650216, China}
\affiliation{Key Laboratory for the Structure and Evolution of Celestial Objects, Chinese Academy of Sciences, Kunming 650216, China}
\email{shenghonggu@ynao.ac.cn}

\author{Xin Liu}
\affiliation{School of Astronomy and Space Science, Nanjing University, Nanjing 210023, China}
\email{liu328@smail.nju.edu.cn}

\author{Xiyun Hou}
\affiliation{School of Astronomy and Space Science, Nanjing University, Nanjing 210023, China}
\email{houxiyun@nju.edu.cn}

\author{Jing Huang}
\affiliation{Yunnan Observatories, Chinese Academy of Sciences, 
Kunming, 650216, China}
\affiliation{University of Chinese Academy of Sciences, Beijing 100049, China}
\email{huangjing@ynao.ac.cn}

\correspondingauthor{Xiaobin Wang}

\begin{abstract}
The near-Earth asteroid (469219) Kamo'oalewa, a quasi-satellite of the Earth, is going to be observed and sampled in situ in 2026-2027 by the Chinese space mission Tianwen-2. Here we analyze Kamo'oalewa's photometric and spectroscopic data to determine its basic physical properties, crucial for the sample return by Tianwen-2. With photometric inversion methods, we derive a spin period of 28.4517~min, an ecliptic pole orientation of $(276^{o}.79, -21^{o}.43)$, and a slightly flattened convex shape. The estimated photometric slope of $0.94^{-0.24}_{+0.18}$~mag/rad suggests a high geometric albedo for Kamo'oalewa, i.e., potentially S-type classification. Using the estimated absolute magnitude of $24.98$~mag, the asteroid's size (diameter) could be 30~m, assuming a typical albedo of S-type asteroids. The spectral classification from a constructed artificial neural network tool also supports the S-type hypothesis: Kamo'oalewa could be a strongly space-weathered fragment of an A-type or Q-type asteroid. Using the derived pole, size, and shape information of the target, we estimated its thermal inertia as $163.0$~Jm$^{-2}$K$^{-1}$s$^{-1/2}$, based on the derived Yarkovsky drift of $A_2=-13.29349563\times10^{-14}$~au/d$^2$. Accordingly, the target could have a surface of a mixture of grains and small boulders, like the surface of asteroid Bennu.
\end{abstract}


\section{Introduction}

Near-Earth asteroid (469219) Kamo'oalewa (2016~HO$_3$) was discovered by the Pan-STARRS telescope at Haleakala on 2016 April 27 and identified as a quasi-satellite of the Earth \citep{DelaFuenteMarcos2016}. As the closest quasi-satellite of the Earth, it is chosen as the target of several missions\citep{Zhang2019,Thirouin2016,Huangetal2020}, for example, the China National Space Administration (CNSA)'s asteroid mission, called Tianwen-2 \citep{Zhanghe2024}. The scientific goal of Tianwen-2 mission is to understand the origin and evolution of small objects of the Solar System. The Tianwen-2 mission is visiting the selected asteroid Kamo'oalewa and the main-belt comet 311P/PANSTARRS. The Tianwen-2 spacecraft was launched on 2025 May 29 and approached Kamo'oalewa around 2026 June 7. According to the mission schedule, the spacecraft will make in situ observations at a 3~km distance, then extract at least $300$~cm$^3$ of sample material from the asteroid's surface. After returning the sample to the Earth, the spacecraft will be directed to another target, the main-belt comet 311P/PANSTARRS.

Until now, due to its faint magnitude, Kamo'oalewa lacks physical characterization. The early observations of Kamo'oalewa \citep{Mastaler2016} with the Pan-STARRS's 1.8-m Ritchey-Chr\'etien telescope provided a magnitude of 21.5 mag in the V-band. \cite{Fohring2017} and \cite{Reddy2017HO3} inferred close synodic periods of 27.90~min and 28.3~min, respectively. \cite{Reddy2017HO3} suggested S-type composition based on spectroscopic data obtained with the Large Binocular Telescope (LBT) using Multi-Object Double Spectrographs (MODS) on 2017 April 14.  \cite{Fohring2017} carried out visible-wavelength spectroscopic and photometric observations using the Keck II telescope in 2018, and found brightness variations of 0.5, 0.4, and 0.2 mag in the V, R, and I band, respectively. Using the photometric data of \cite{Reddy2017HO3}, \cite{Li2021} estimated a possible ellipsoidal shape of the Kamo'oalewa of $b/a=0.48$. \cite{Sharkey2021} suggested an S- or L-type classification, proposing that Kamo'oalewa had undergone extensive space weathering or comprised lunar materials, based on the absorption feature and red slope in its spectrum. 

The paper aims to determine the physical properties of Kamo'oalewa. We derive its convex shape, spin parameters, absolute magnitude, and photometry slope with two photometric inversion methods \citep{Kaasalainen2001a, Muinonen2020} in Section 2. With a constructed artificial neural network (ANN) tool, the VNIR reflectance spectrum of Kamo'oalewa from the literature \citep{Sharkey2021} is re-analyzed in Section 3. Using the derived pole solution and size of Kamo'oalewa, the thermal inertia is estimated in Section 4, based on the derived Yarkovsky drift. The summary on the present study is given in the last section.

\section{Photometric analysis} \label{sec:photometry}

We collected photometric data of the Kamo'oalewa from the MPC database \footnote{\url{https://www.minorplanetcenter.net//db_search/}}, including 17 nights of data obtained between 2016 and 2018. The observational information is listed in Table~\ref{observations}, which includes start time of each lightcurve, heliocentric and geocentric distances of the asteroid, solar phase angle, ecliptic coordinates of the Kamo'oalewa in J2000 corresponding to that time, the filter, the observatory code of the used telescope and data point number. Among the 17 nights of photometric data, 14 nights of data were obtained through the G-band filter and the rest through the R-band filter.  
Before the photometric inversion, all involved photometric data were transferred to the V-band according to the color indexes V-R=0.37 and V-G=0.14, and into the reduced magnitudes (magnitudes for the asteroid at 1~au distance to the Sun and observer). The light travel time of each observation was corrected according to the distance from the asteroid to the observer. The input data of Kamo'oalewa used in photometry inversions mainly distributed at 4 different apparitions, the phase angle spanned a range from  43 to 73 degrees, and its apparent magnitude varied between 24.7 and 27.6 mag. 

\begin{table*}
\caption{Information of used observational data.}
\label{tab:observation}
\centering  
\begin{tabular}{l l l l l l l l}    
\hline       
Date    & r & $\Delta$ & Phase & (Long., Lati.) &Filter &Code&Pts \\ 
    UTC     &  AU  & AU &Degree      & J2000  &   & &\\
\hline    
2016  5  10.29&  1.087&  0.135&  52.15& (225.00, 2.87)&R&568&3\\
2016  5  11.35&  1.087&  0.135&  52.15& (225.00, 2.87)&R&568&4\\
2016  5  18.30&  1.080&  0.140&  57.50& (231.21, 2.06)&R&568&10\\
2016  6  10.28&  1.050&  0.161&  73.39& (250.88,-0.60)&R&568&12\\
2017  3  23.55&  1.104&  0.170&  47.43& (185.17, 6.83)&G&T12&10\\
2017  3  25.57&  1.104&  0.167&  46.75& (186.81, 6.72)&G&T12&10\\
2017  3  29.52&  1.105&  0.161&  45.47& (190.00, 6.49)&G&T12&10\\
2017  3  30.53&  1.105&  0.160&  45.17& (190.81, 6.43)&G&T12&10\\
2017  3  31.53&  1.105&  0.158&  44.88& (191.61, 6.37)&G&T12&10\\
2017  4   1.56&  1.105&  0.157&  44.59& (192.45, 6.30)&G&T12&11\\
2017  4  26.42&  1.098&  0.139&  45.78& (212.48, 4.36)&G&T12&35\\
2017  4  27.43&  1.098&  0.139&  46.21& (213.30, 4.27)&G&T12&35\\
2017  4  28.43&  1.097&  0.139&  46.65& (214.11, 4.18)&G&T12&25\\
2017  5  25.39&  1.072&  0.155&  63.65& (238.02, 1.14)&G&T12&20\\
2017  5  28.31&  1.068&  0.157&  65.54&(238.84, 1.03)&G&T12&11\\
2017  5  29.28&  1.068&  0.157&  65.54&(238.84, 1.03)&G&T12&11\\
2018  4  13.50&  1.104&  0.148&  43.98& (201.26, 5.52)&G&T12&20\\
\hline    
\label{observations}
\end{tabular}

Notes:'568'--the 3.6-m Canada-France-Hawaii Telescope at Mauna Kea, 'T12'--the University of Hawaii 88-inch telescope at Maunak.
\end{table*}

\subsection{Photometric inversion methods}
The photometric inversion aims to determine the asteroid's physical parameters by comparing certain a disk-integrated brightness model to the observed values. The brightness model often contains information of shape, size, spin status, and scattering properties of an asteroid and observational geometry. In practice, the brightness of asteroid at a certain observational geometry (representing observer and source directions with $E,E_0$) is calculated by summing radiation reflected by the illuminated and visible surface area: 
\begin{equation}
\begin{aligned}
   L(E,E_0)=\Sigma F*S(\mu_i,\mu_{0,i})G(\vartheta_i,\psi_i)\sigma_i,\\
   \mu=E\cdot n, \mu_0=E_0\cdot n. 
\label{brightness}
\end{aligned}
\end{equation}
Here, $F$ is the inducing intensity of solar light on surface of asteroid. $\mu_0$ and $\mu$ are the inner products of the normal vector of a facet of convex shape with vectors of the light source and the line of sight, respectively. The number of facet depends on the scheme for triangulation of a spherical surface. $S(\mu_i,\mu_{0,i})$ is the so-called scattering law, the size of the $i$-th facet is computed by its Gaussian density $G(\vartheta_i,\psi_i)$, which is approximated with truncated spherical harmonics in the practical computation.

As comparison, two photometry inversion methods are applied to Kamo'oalewa's photometric data. The first one was developed by \cite{Muinonen2015} and \cite{Muinonen2020, Muinonen2022}, hereafter called as M-inversion, another by \cite{Kaasalainen2001a} and \cite{Kaasalainen2001b}, called as K-inversion. The main differences of the two inversion methods are (1) the scattering law functions used in the brightness model of the asteroid, and (2) the algorithms used to derive the parameter solutions through lightcurve inversion. 

The brightness model of M-inversion uses the Lommel-Seeliger law and a phase function of $H,\!G_1,\!G_2$ (Eq.~\ref{M_S function}), whereas a combination of Lambert and Lommel-Seeliger laws with a linear-exponential phase function (Eq.~\ref{k_S function}) is applied in the K-inversion, 
\begin{equation}
\begin{aligned}
S(\mu_0,\mu,\alpha) =2p_v\frac{\Phi_{HG_1G_2(\alpha)}}{\Phi_{LS}(\alpha)}\frac{\mu \mu_0}{\mu+\mu_{0}}.
\end{aligned} 
\label{M_S function}
\end{equation}
Here, $\alpha$ is the solar phase angle, $\Phi_{LS}(\alpha)$ is the Lommel-Seeliger phase function (details see Eq.(6) of \cite{Muinonen2020}), $\Phi_{HG_1G_2(\alpha)}$ is the phase function of the three-parameter $H,G_1,G_2$ magnitude system \citep{Muinonen2010}, and $p_v$ the geometric albedo. $G_1,G_2$ are the phase function parameters in the M-inversion. Considering the phase angle distribution of Kamo'oalewa, we actually fit the photometric slope formula (details see Eq.(12) in \cite{Muinonen2020}) instead of the $H,G_1,G_2$ formula.

\begin{equation}
\begin{aligned}
S(\mu,\mu_{0})=f(\alpha)\frac{\mu \mu_0}{\mu+\mu_{0}}(1+c),\\
f(\alpha) = A_0 e^{\frac{-\alpha}{D}}+k\alpha+1.
\end{aligned}
\label{k_S function}
\end{equation}
Here $A_0$, $D$ and $K$ are the phase function parameters in the K-inversion. 

The M-inversion applies both a least-squares algorithm and a MCMC method. During the downhill least-squares fitting, we firstly estimated the involved parameters by setting the same weight for each lightcurve in the chi-square computation, then the weighted chi-square (see Eq.(\ref{chi-weights})) was used in the re-fitting procedure:
\begin{equation}
\begin{aligned}
    \chi^{2}(\mathbf{P}) = & \sum_{k=1}^{K} \frac{(-2.5 \mathrm{lge})^2}{\sigma^2_{\epsilon,k}} 
    \times \sum_{j=1}^{N_{k}} \left[\frac{L_{\mathrm{obs},kj}}{L_k} -\frac{L_{\mathrm{mod},kj}(\mathbf{P}) }{L_{\mathrm{mod},k}}\right]^2,\\
    \sigma_{\epsilon,k}=& \sqrt{\frac{N_{k}}{N_{k,\mathrm{eff}}}} \max(\sigma_{0,k},\sigma_{pr,k}). \quad 
\label{chi-weights}    
\end{aligned}
\end{equation}
Here, the $L_{\mathrm{obs},kj}$ and $L_{mod,kj}$ are observed and computed brightnesses. The weight $1/\sigma^2_{\epsilon,k}$ depends on the efficient observational number $N_{k,\mathrm{eff}}$ of each lightcurve (see Eq.(37) in the paper of \cite{Muinonen2020}). The $\sigma_{pr,ik}$ is a prior threshold of uncertainty of the lightcurve, and the $\sigma_{0,k}$ is the root mean square (RMS) of the modeled lightcurve fitted to the observed one. In this work, $\sigma_{pr,ik}$ is set as 0.005 mag for dense lightcurves and 0.02 mag for loose absolute photometric data. 

After some rounds of iterative least-square fitting, we can obtain the final least-squares solutions for both M-inversion and K-inversion procedures. Furthermore, the M-conversion applies a MCMC procedure starting from its derived least-squares solution. The MCMC procedure contains a virtual-observation MC sampler and MCMC sampler. In the MC sampling procedure, the proposal probability density functions (PDFs) of involved parameters are composed of the virtual least-squares solutions, derived from the virtual observations (real data added Gaussian random quantities of given uncertainties and a covariance matrix). Using the proposal PDF, the MCMC sampling was performed to derive the posterior distributions of involved parameters, from which the best values of involved parameters and their uncertainties could be estimated. 

To obtain the solution of involved parameters of an asteroid, the K-inversion applies the Levenberg–Marquardt algorithm based on reduced chi-square value: 
\begin{equation}
\chi^2(\mathbf{P})=\sum_i\sum_j\left[\frac{(L_{obs,i,j}}{L_j}-\frac{L_{mod,i,j}(\mathbf{P})}{L_{mod,j}}\right]^2.
\label{k_chi}
\end {equation}
Here $L_j$ and $L_{mod,j}$ are the averaged observed and modeled fluxes of the $j-th$ lightcurve.

Shortly, the involved parameters in the photometry inversion methods are the shape parameters represented with spherical harmonics coefficients, spin parameters (pole orientation $\lambda,\beta$, spin period $per$, and initial spin phase $\phi_0$) and phase function parameters.  

\subsection{Result of the M-inversion}
\begin{figure*}[ht!]
\centering
\includegraphics[width=1.0\columnwidth]{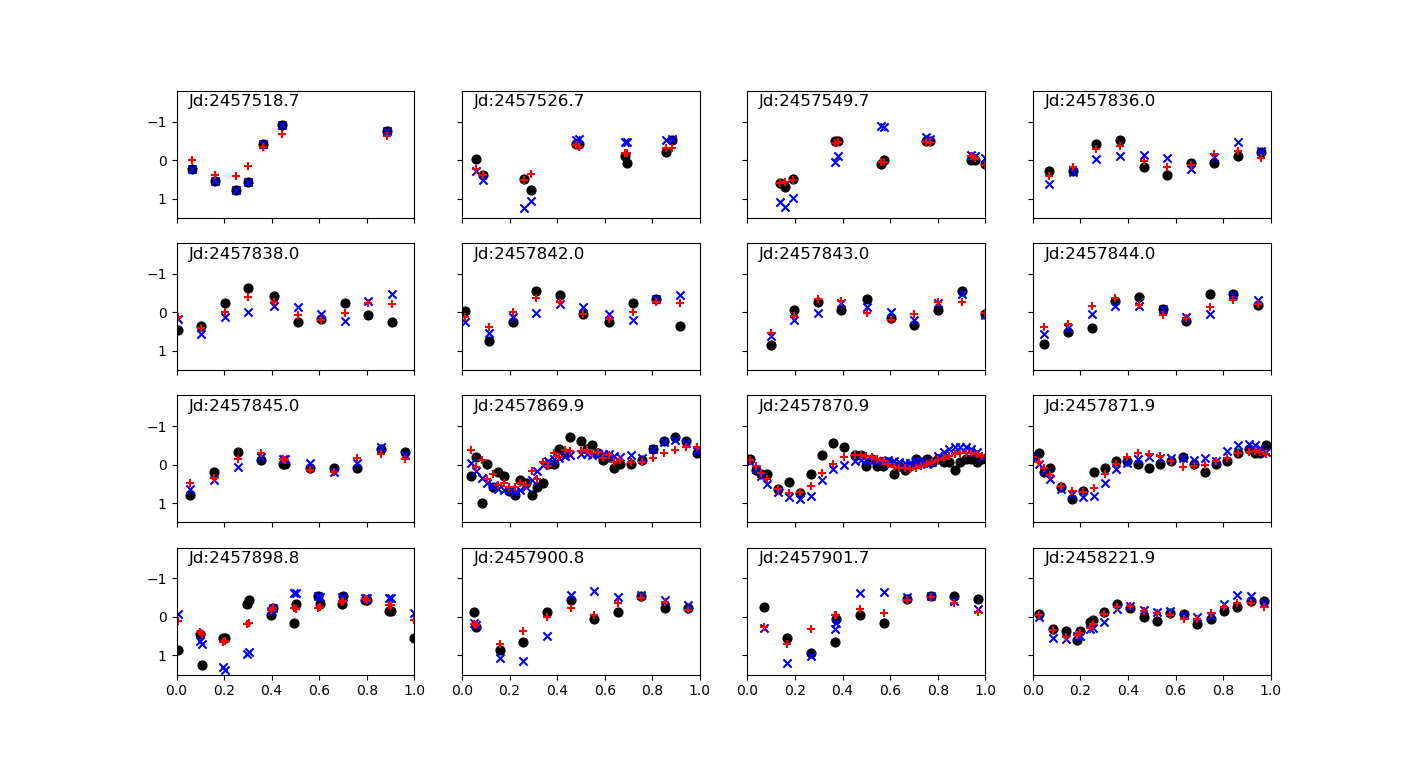}
\caption{The photometric data (black solid dots) of Kamo'oalewa, and the least-squares models by the M-inversion (red pluses) and K-inversion (blue crosses).}
\label{lcskm}
\end{figure*}

\begin{figure*}[ht!]
\begin{center}
\includegraphics[width=0.33\columnwidth]{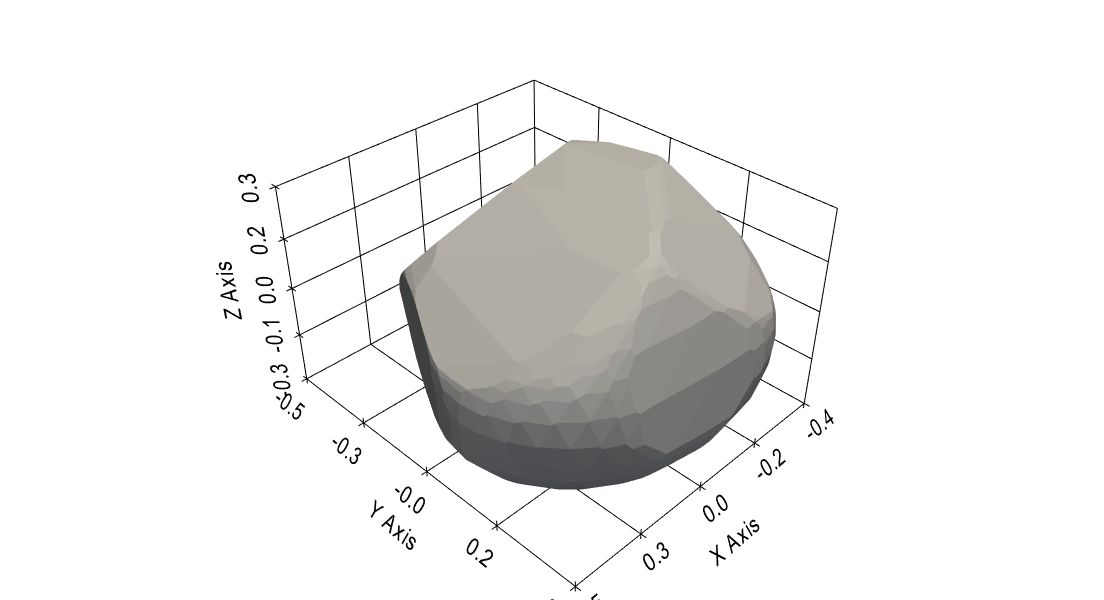}
\includegraphics[width=0.33\columnwidth]{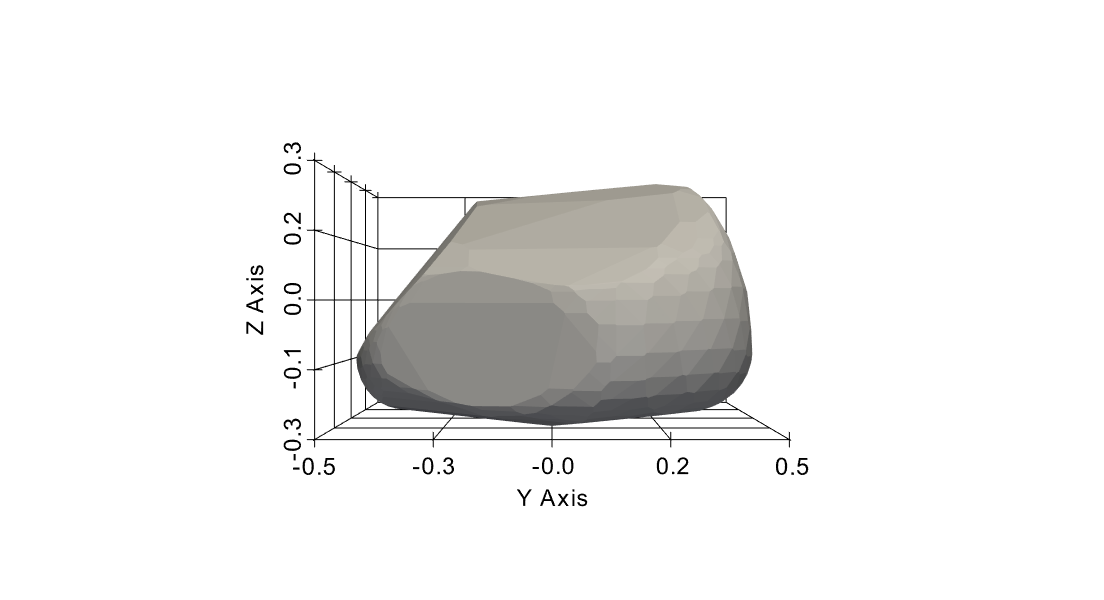}
\includegraphics[width=0.33\columnwidth]{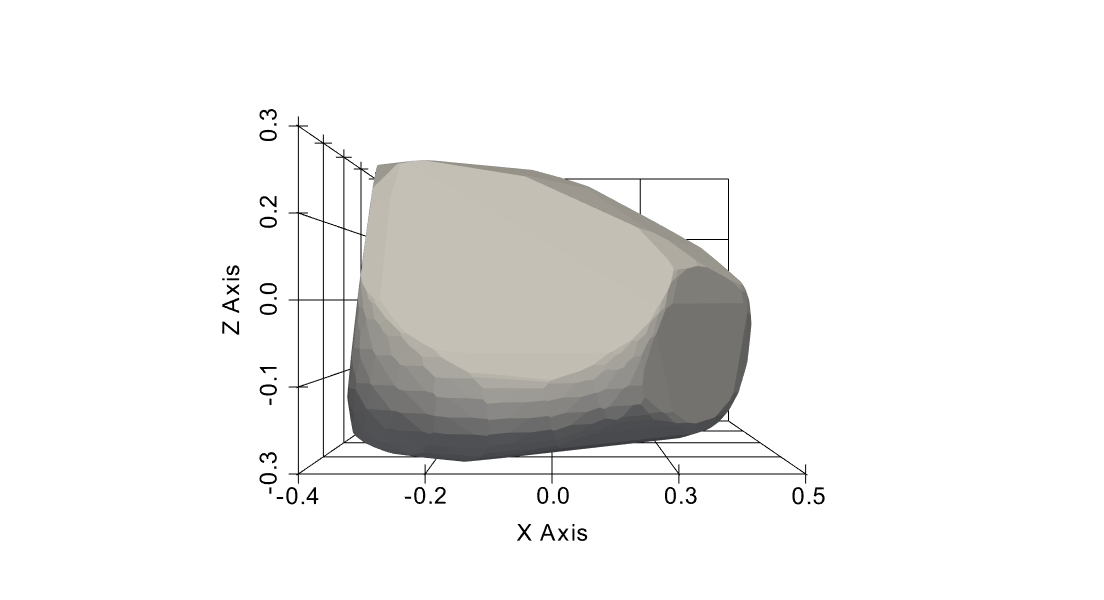}
\includegraphics[width=0.33\columnwidth]{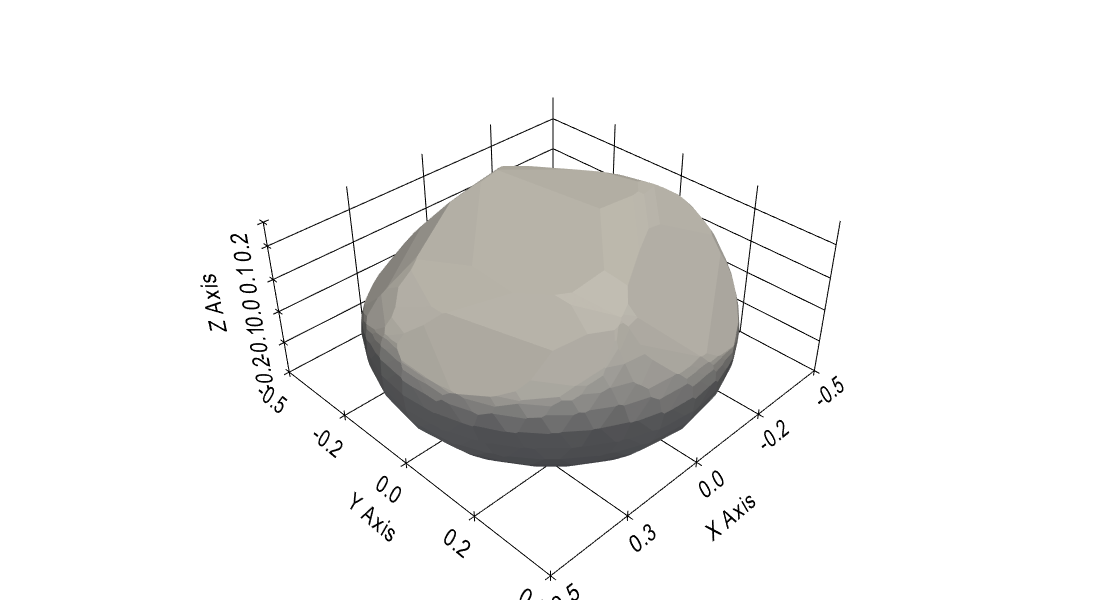}
\includegraphics[width=0.33\columnwidth]{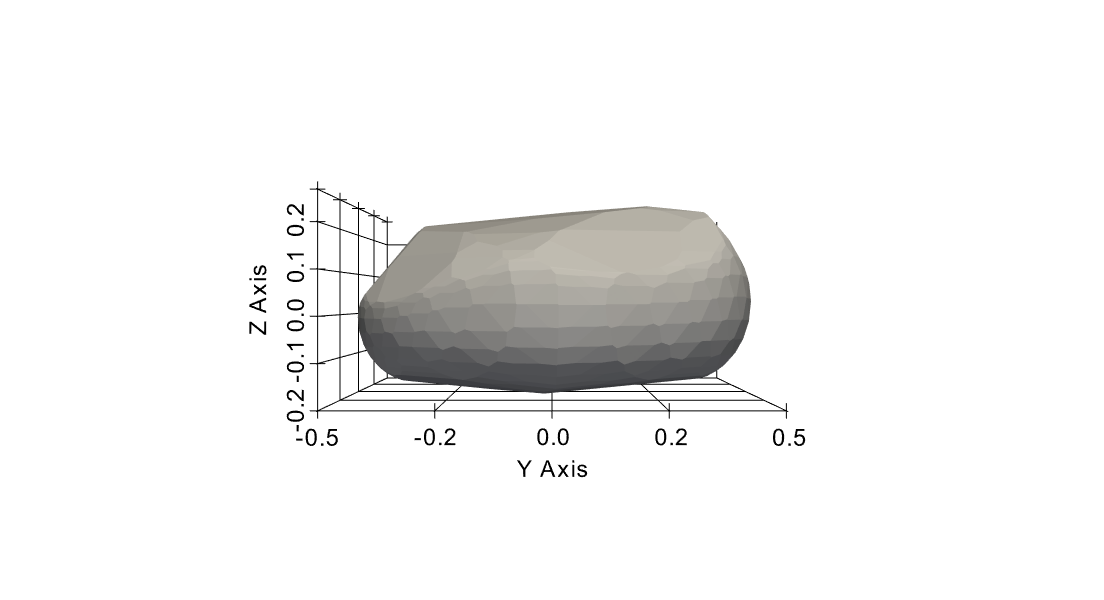}
\includegraphics[width=0.33\columnwidth]{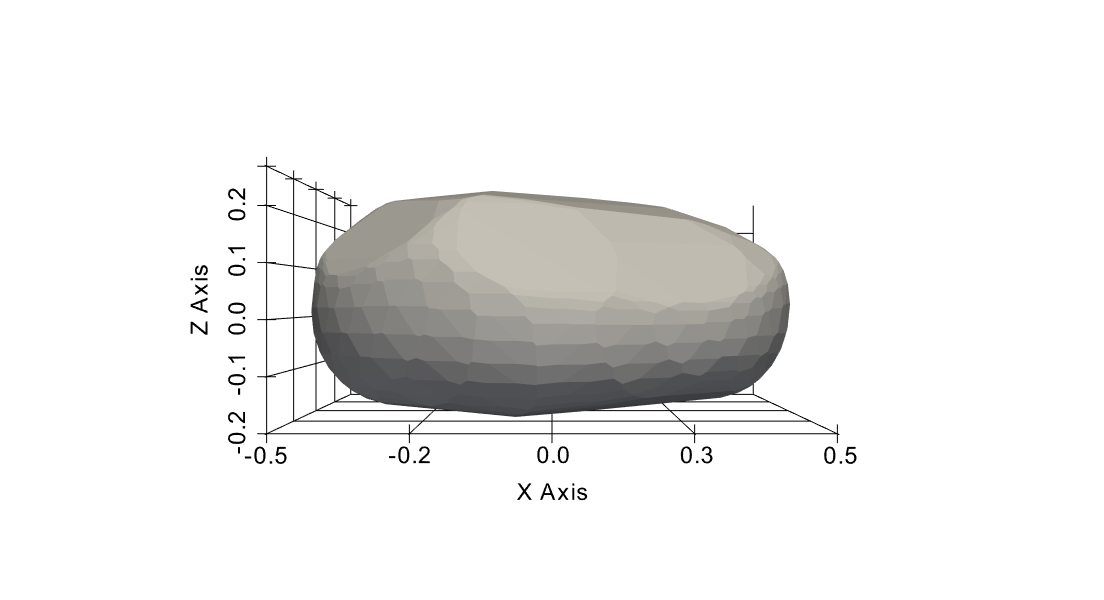}
\includegraphics[width=0.33\columnwidth]{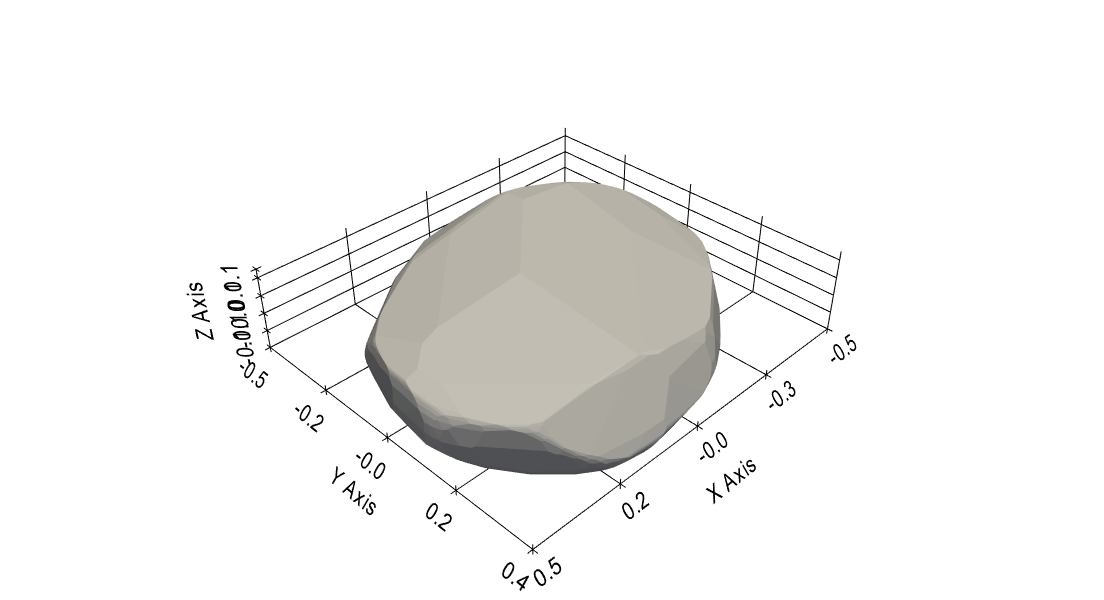}
\includegraphics[width=0.33\columnwidth]{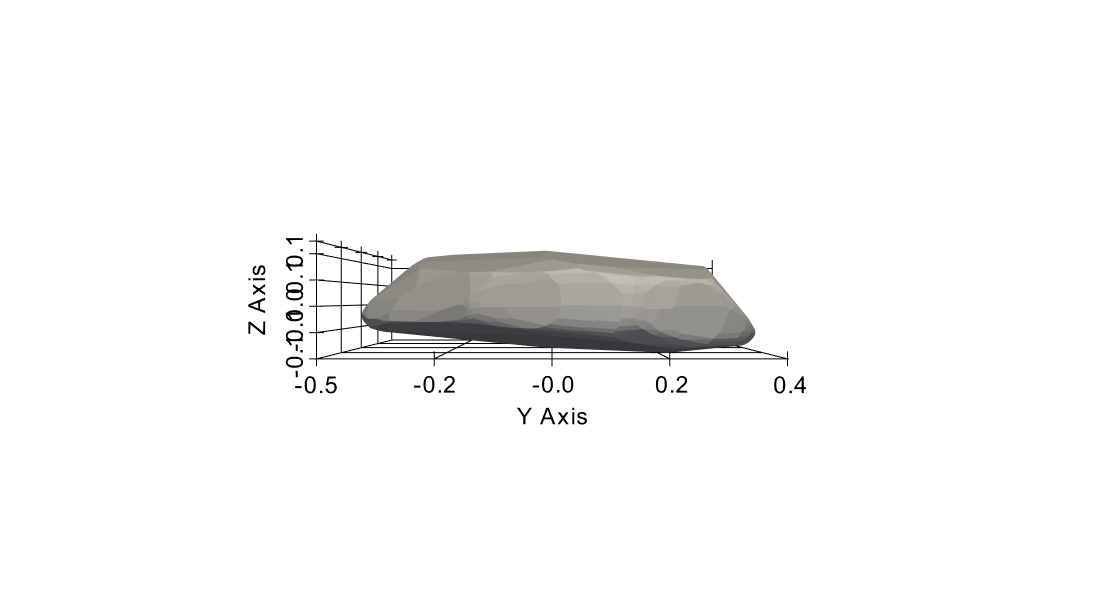}
\includegraphics[width=0.33\columnwidth]{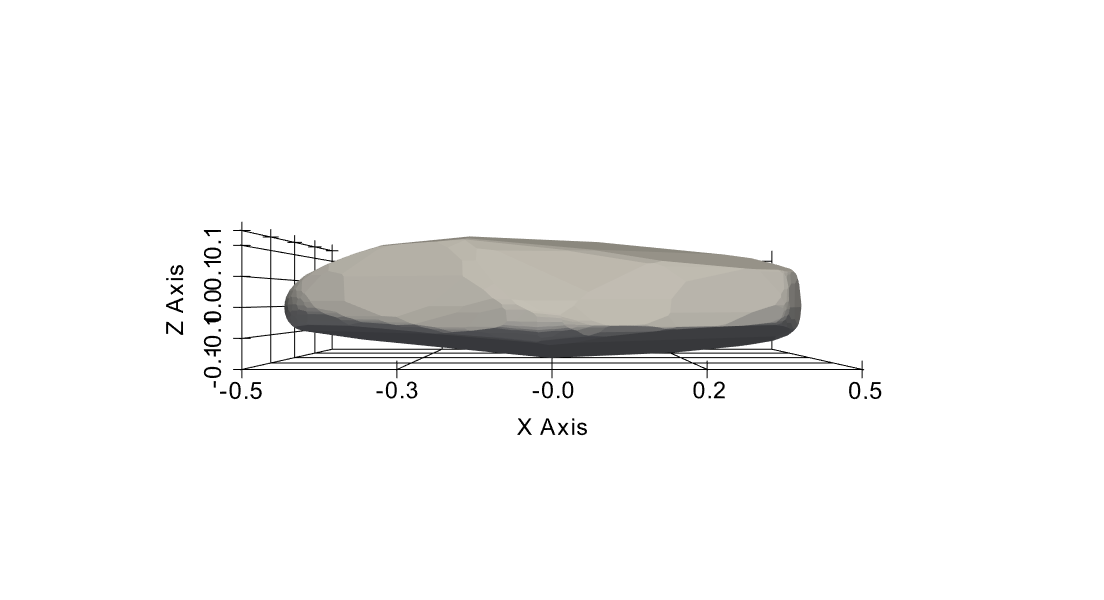}
\end{center}
\caption{The upper and middle rows are the convex shapes of Kamo'oalewa derived by the M-inversion from the combined data set and relative data set, the bottom row from the K-inversion.}
\label{shapes_HO3}
\end{figure*}

In the M-inversion, we used two input data sets: one was composed of 17 nights of relative lightcurves; another one contained 29 absolute photometric data points besides the previous relative data set. Taking the spin period of \cite{Reddy2017HO3} (28.3 minutes) as the initial spin period, we did a systematic scan over the unit sphere with a 10-degree mesh step to search the initial value of Kamo'oalewa's pole. Then the Levenberg–Marquardt algorithm was run iteratively to minimize the weighted chi-square (see Eq.(\ref{chi-weights})). In this step, a 18-rows triangulation per octant and a maximum degree of spherical harmonics $l=m=8$ were set. For the case of of relative data set, we derived the least-squares solution for spin parameters as  $(277^{o}.12,-16^{o}.31)$ in the ecliptic coordinate with a spin period of 28.451699 minutes; whereas, for the case of combined data set, the pole solution is $(276^{o}.79,-21^{o}.43)$ with a spin period of 28.45170 minutes. Because of the lack of data at small solar phase angles, we fitted the photometry slope and derived a value of 0.998~mag/rad for its photometry slope with the combined data set. Figure \ref{lcskm} shows the least-squares fitting (red pluses) to the observed lightcurves of Kamo'owlewa. With the derived spherical harmonics coefficients, the sizes of facets were computed with their corresponding Gaussian surface densities, then the convex shapes were estimated by a Minkoski optimization procedure.The first row of Fig.\ref{shapes_HO3} is the shape solution from the combined data set, and the middle row is that from the relative data set. The ellipsoid fits to the two convex shapes are $a:b:c=1:0.96:0.53$ and $a:b:c=1:0.89:0.67$, respectively.

\begin{figure*}[ht!]
\begin{center}
\includegraphics[width=0.45\columnwidth]{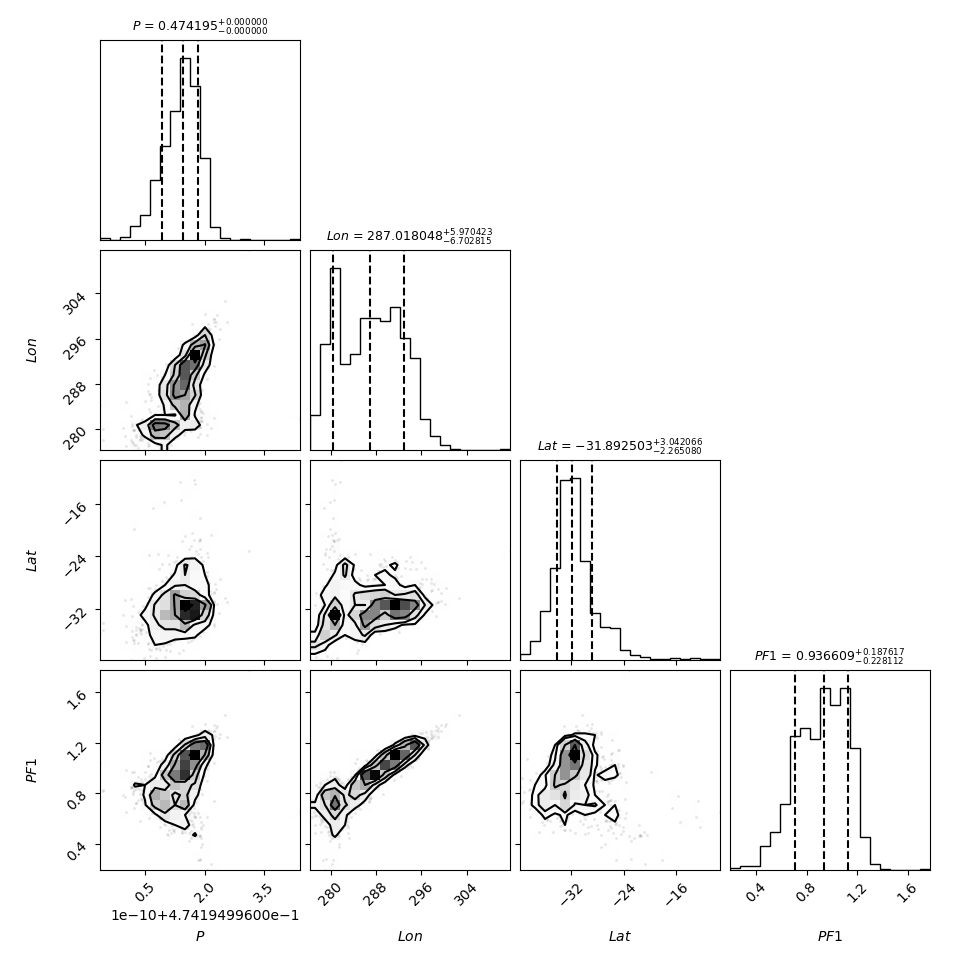}
\includegraphics[width=0.40\columnwidth]{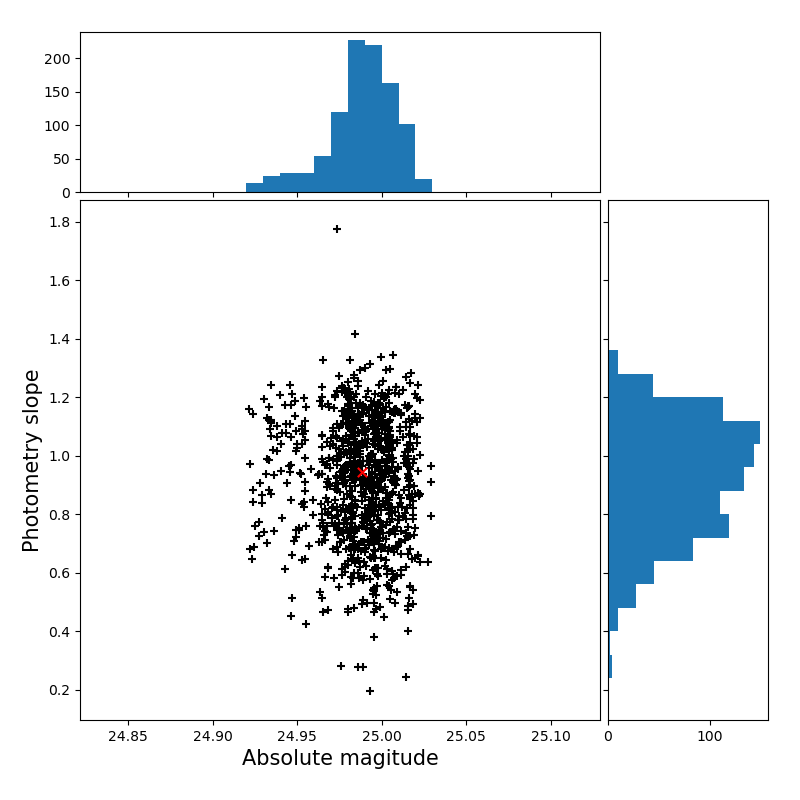}
\end{center}
\caption{Left: The posterior distributions of spin parameters. Right: The posterior distributions of absolute magnitude and photometry slope. The red cross marks the best value.}
\label{spin-par}
\end{figure*}

With the help of virtual-observation MCMC sampler procedure, we can estimate the uncertainties of the parameters derived above. During this procedure, a 18-rows triangulation per octant and a maximum degree of spherical harmonics was set at $l=6$. The MCMC analysis procedure started from a set of 1000 virtual least-squares solutions of virtual-observation data. Then a random-walk MCMC sampling was performed from the proposal PDF and posterior distributions of involved parameters were derived. Figure~\ref{spin-par} shows the posterior distributions of spin parameters (left) and those of the photometric slope and absolute magnitude (right). The peak values of pole and period are around $(287^{o}.02^{-6.70}_{+5.97}, -31^{o}.89^{-2.27}_{+3.04})$, and $28.4517\pm2.0\times 10^{-7}$ minutes. The absolute magnitude $H$ and photometric slope $\beta$ are $24.98^{-0.17}_{+0.18}$ mag and $0.94^{-0.24}_{+0.18}$, respectively. Referring to the recent work of \cite{Xuetal2025} and \cite{Pentikainen2026}, the small photometric slope value of Kawo'oelewa suggests consistency with S-type composition. Assuming a geometric albedo of 0.24, the equivalent diameter of Kawo'oelewa could be $\sim30$ meters.

\subsection{Result of the K-inversion}
In the photometric inversion to the Kamo'oalewa data with the K-inversion, a 4-row triangulation per octant and a maximum degree of spherical harmonics $l=8$ were set. In this procedure, the 17 nights of relative photometric data were included and the related scattering parameters $(a, d, k,$ and $c)$  were fixed (in detail, $a=0.5, d=0.1, k=-0.5, c=0.1$). Thus, only the shape parameters and spin parameters were estimated. With a least-square algorithm, we derived the pole orientation as $(273.^{o}58, -20^{o}.26)$ in the ecliptic coordinate with a spin period of 28.451704 minutes, and a flatter convex shape (see the bottom row of Fig.\ref{shapes_HO3}). This pole solution is very close to that from the M-inversion, the ellipsoid fitting to this convex shape is $a:b:c=1:0.80:0.52$.

\begin{table}
\caption{Parameters of Kamo'oalewa derived by the least-squares algorithms.}
\label{tab:photometry_inversion}
\centering  
\begin{tabular}{llll}    
\hline
Parameter&M-inversion& M-inversion&K-inversion\\
&Combined data set& Relative data set& Relative data set\\ 
\hline\hline
Pole & (276.79,-21.43) & (277.12,-16.31)&(273.58,-20.26) \\
Period& 28.451700 minutes& 28.451699 minutes &28.451704 minutes \\
Ellipsoid fit of convex shape&1:0.89:0.67&1:0.96:0.53&1:0.80:0.52\\
Photometry slope&0.998 & - & - \\
\hline
\label{spin parameters}
\end{tabular}
\end{table}

\section{Taxonomy analysis}
The compositional information of Kamo'oalewa is important for figuring out its origin and evolution, as well as for the sample return task of the Tianwen-2 mission. But the spectroscopic data available for Kamo'oalewa is scarce. Here, we used the only VNIR spectrum from the literature \citep{Reddy2017HO3,Fohring2017}. This spectrum was composed of the visual spectrum obtained by the KECK II telescope in 2018 and the near-infrared spectrum obtained with the LBT on 2017 April 14. In the near future, the payload instruments of Tianwen-2, i.e., the multi-spectral camera \citep{Scientia2025-279506} and visible-infrared imaging spectrometer \citep{Scientia2025-279507}, will provide plenty of spectroscopic data. For the analysis on the existing and forthcoming spectral data of Kamo'oalewa, we established an ANN tool. In the following, we present the trained ANN and the results of the analysis for the spectroscopic data available.

\subsection{The ANN tool}
The constructed ANN model is composed of 3 layers: one input layer of 72 neurons (the number of features of the input spectrum), one hidden layer of 30 neurons, actually a full connection layer, and one output layer of 8 neurons (8 involved asteroid types: $S$, $S_a$, $S_q$, $S_r$, $V$, $O$, $Q$, and $R$). To training the ANN models, we used the SMASS II data \citep{DeMeo2009} and the MIT-Hawaii Near-Earth Object Spectroscopic Survey data \citep[MITHNEOS,][]{Binzel2019}. The referred labels of selected samples are from the Bus–DeMeo taxonomy system \citep{DeMeo2009}. In practice, 178 samples with taxonomy labels were selected (the detailed information is listed in Table \ref{samples}). The training data were trimmed according to the wavelength range (0.40-1.25 microns) of the available spectral data of Kamo'oalewa. 

With a similar procedure described in the paper of \cite{Luo2024}, several tens of ANN models were trained. The detailed procedures include (1) re-sampling these involved training data according to the resolution of available spectrum data of Kamo'oalewa; (2)  cloning synthetic spectra to balance the sample number among involved 8 types of asteroids; and (3) training an ANN model by around 1000 epochs with a learning rate of 0.001. The training data set composed of 800 pieces of spectral data (real plus cloned), were divided into a training data set and a test data set according to a ratio of 8:1 in each epoch. The training procedure applied the MBGD (Mini-Batch Gradient Descent) algorithm. That is, for each epoch the mini-batch training data (here we set 32 samples) are randomly selected from the training data set. In practice, several tens of ANNs were trained. The final taxonomy result for the tested asteroid spectra, was determined through voting results of multiple ANN models. Using the predicted and labeled types of 178 real spectra, the accuracy of the ANN models built was estimated, the averaged accuracy was 0.81. The detailed accuracy for each class is listed on the second row of Table \ref{samples}. 

\begin{table}
\caption{Number of samples for each taxonomic type in our study.}
\label{tab:samples}
\centering  
\begin{tabular}{c c c c c c c c c}    
\hline\hline       
Class & S & $S_a$ & $S_q$ & $S_r$ & V & $S_{qw}$ & Q & R  \\ 
\hline
$Number$ & 86 & 17 & 32 & 8 & 20 & 9 & 4  & 2  \\
\hline
 Accuracy& 0.73 & 0.94 & 0.88 & 0.75 & 0.95 & 0.56 & 0.50 & 1.0 \\
\hline 
\label{samples}
\end{tabular}
\end{table}
%

\begin{figure*}[ht!]
\begin{center}
\includegraphics[width=0.22\columnwidth]{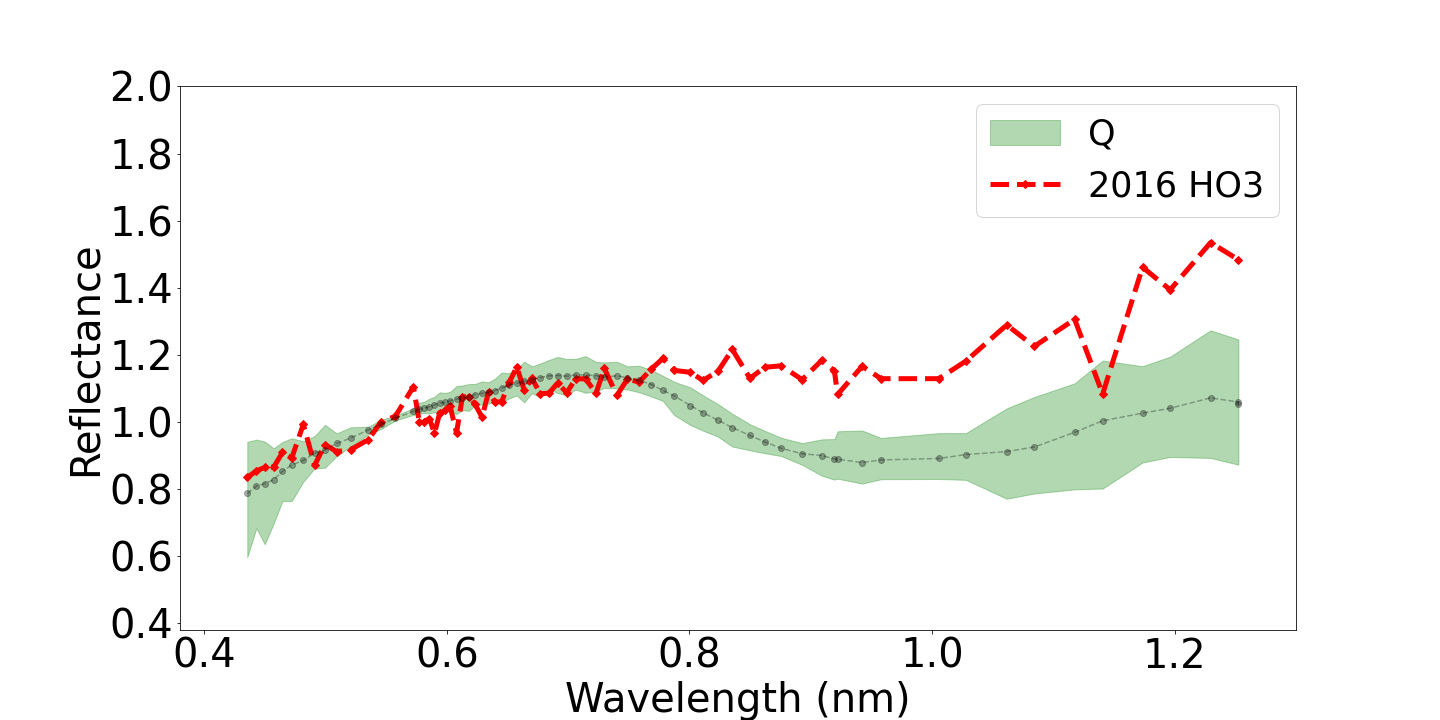}
\includegraphics[width=0.22\columnwidth]{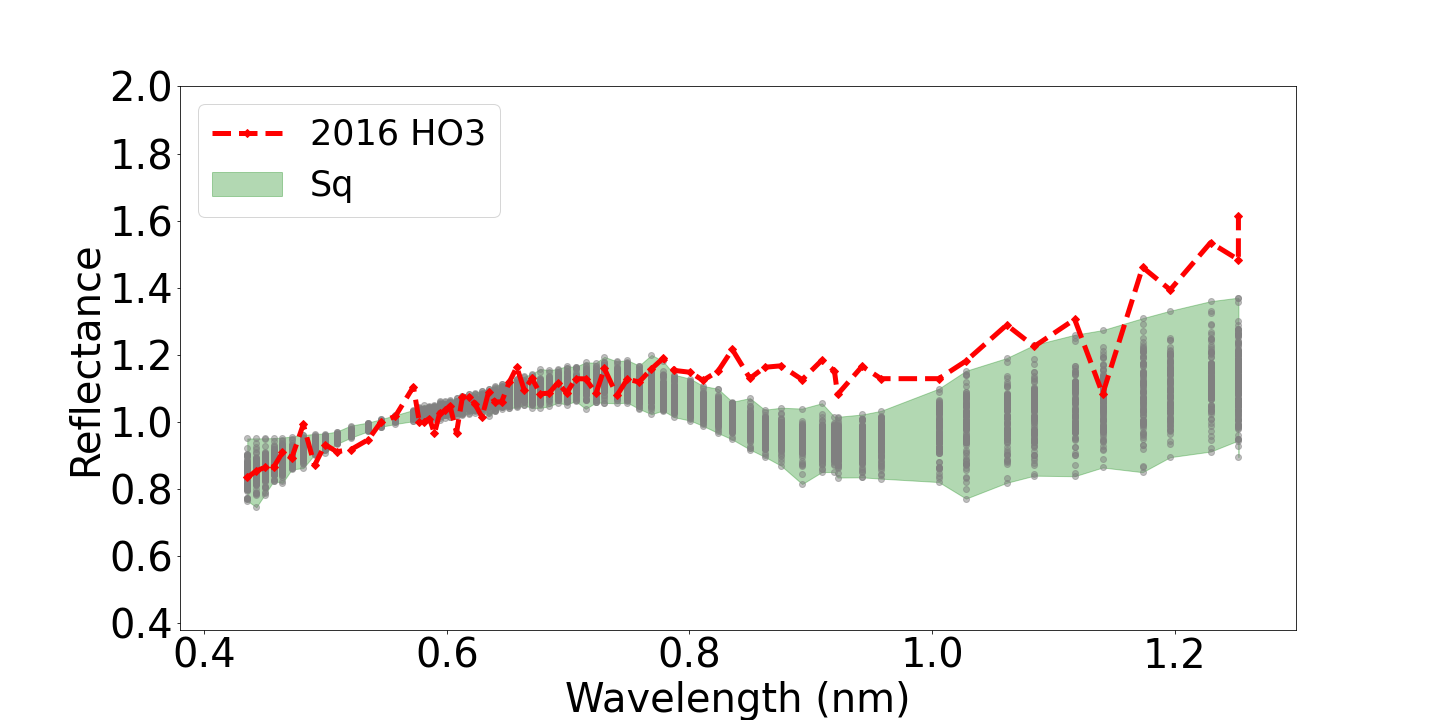}
\includegraphics[width=0.22\columnwidth]{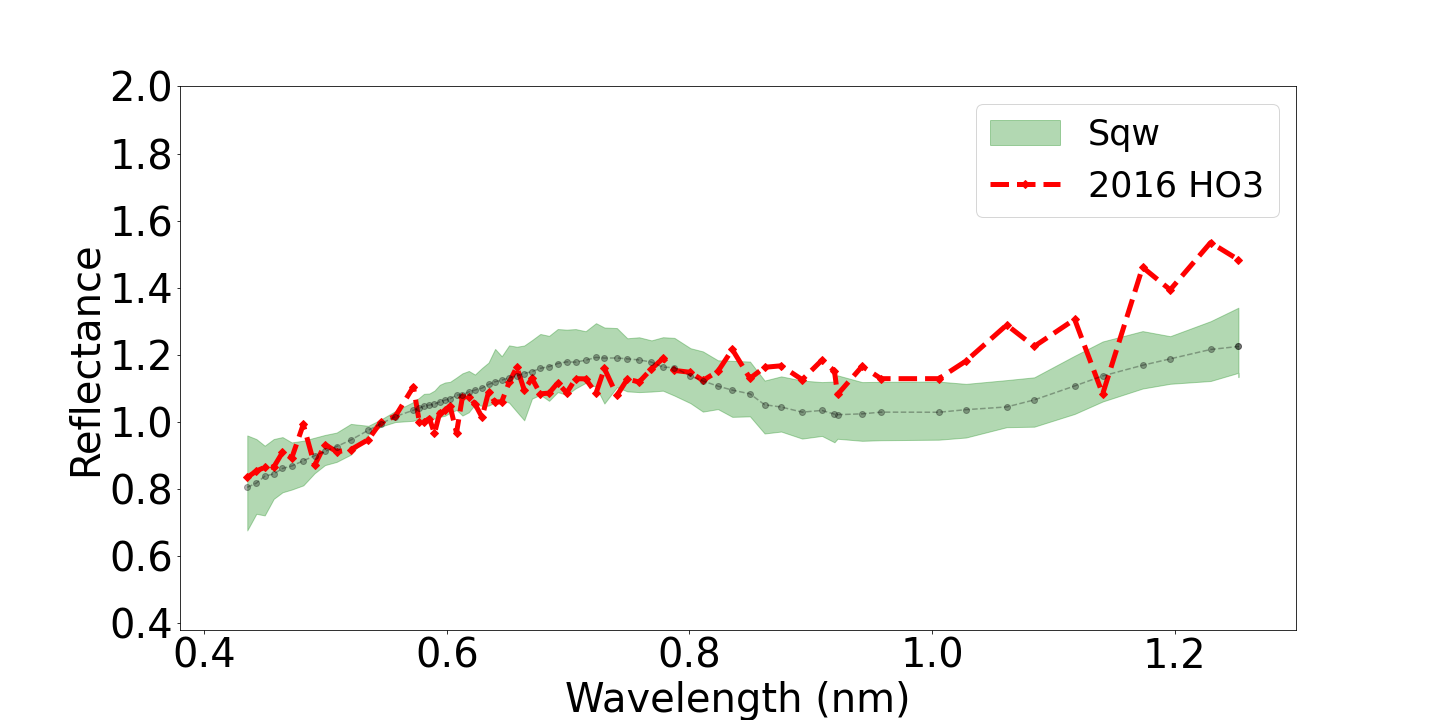}
\includegraphics[width=0.22\columnwidth]{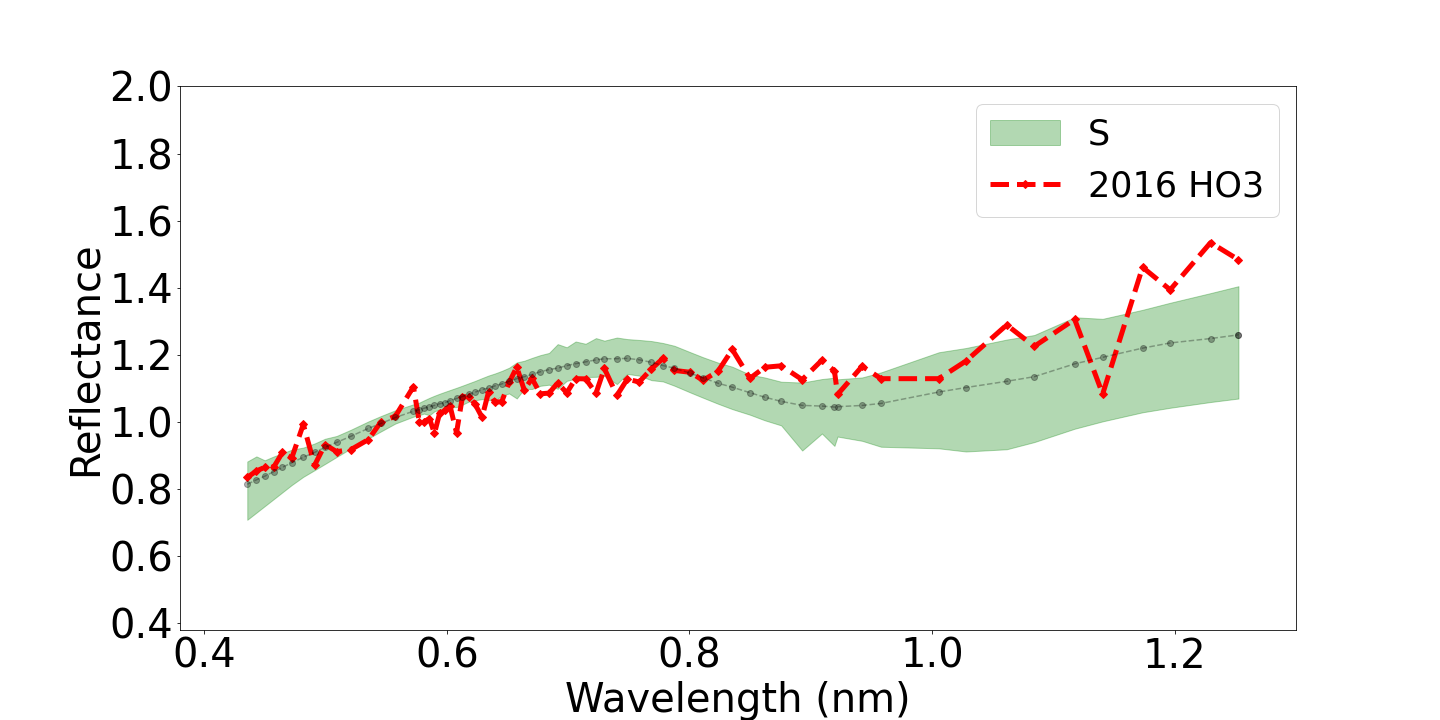}
\includegraphics[width=0.22\columnwidth]{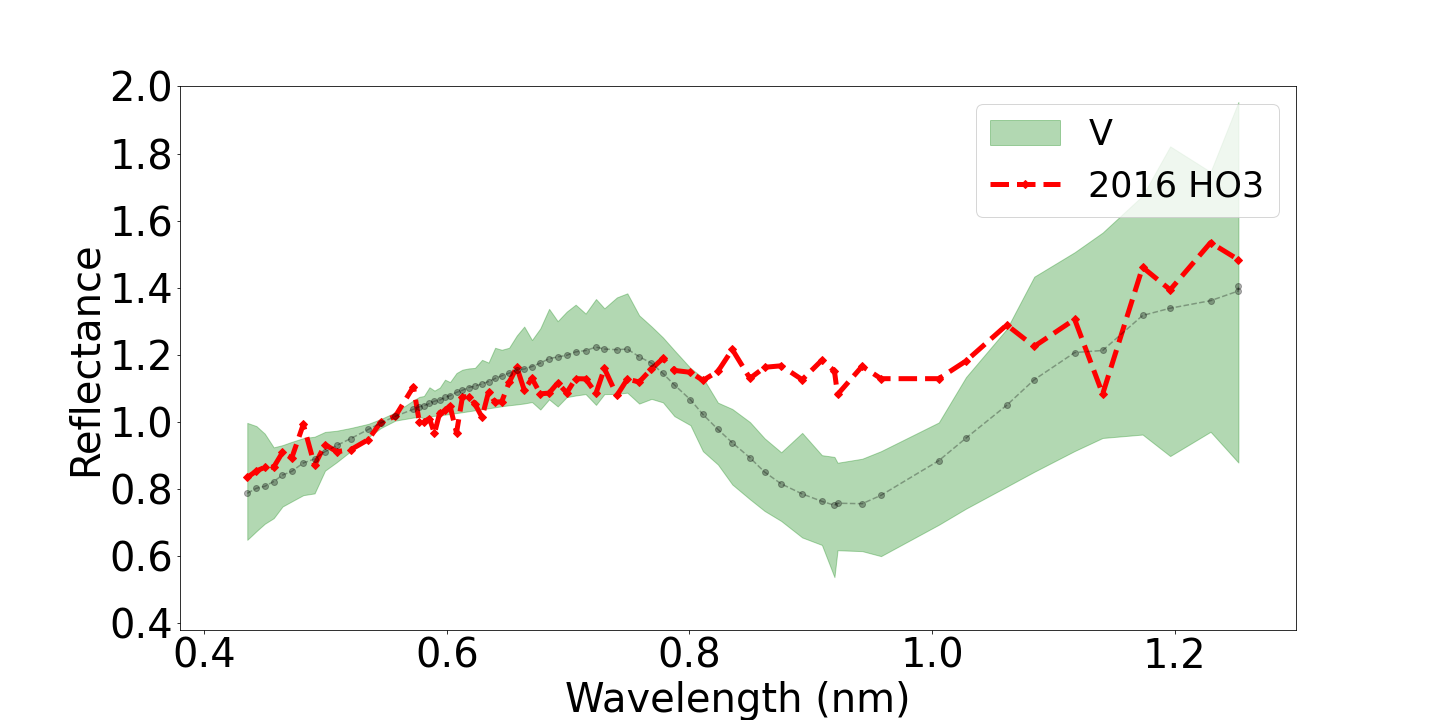}
\includegraphics[width=0.22\columnwidth]{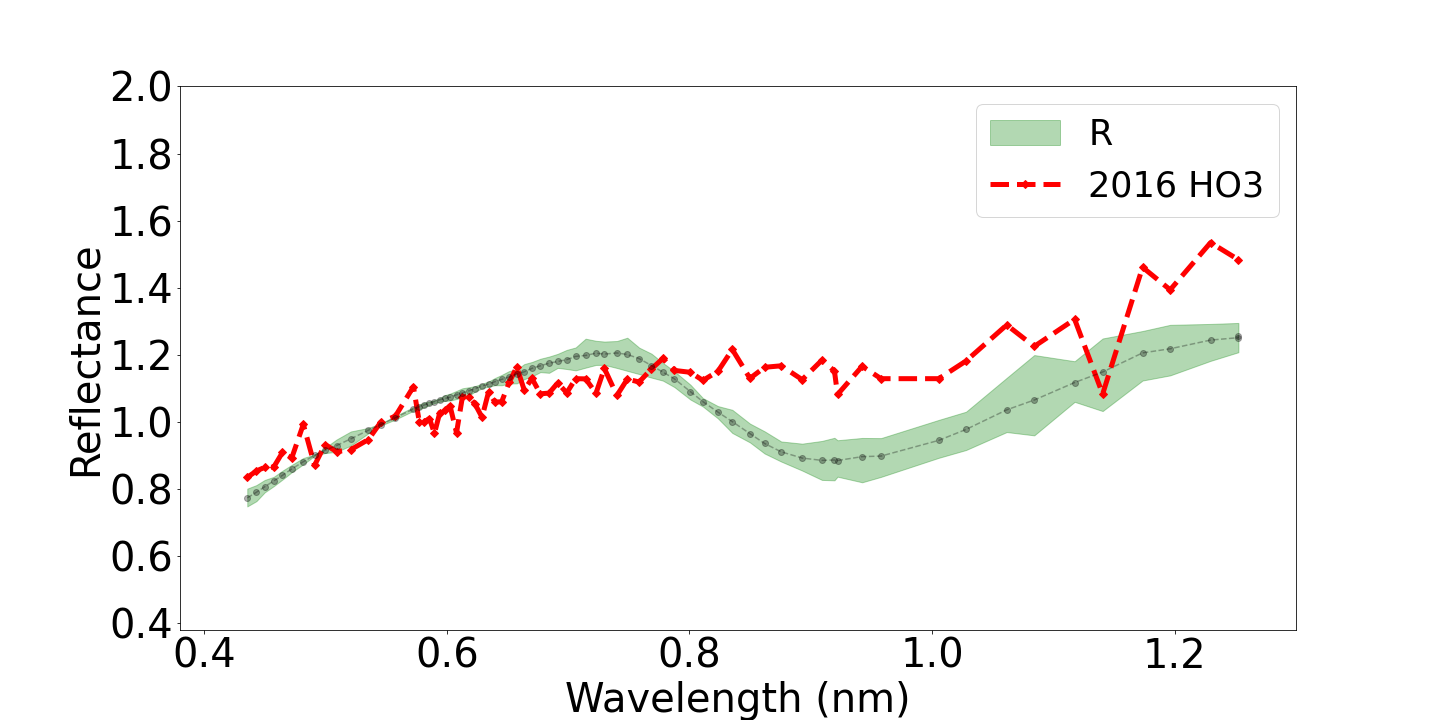}
\includegraphics[width=0.22\columnwidth]{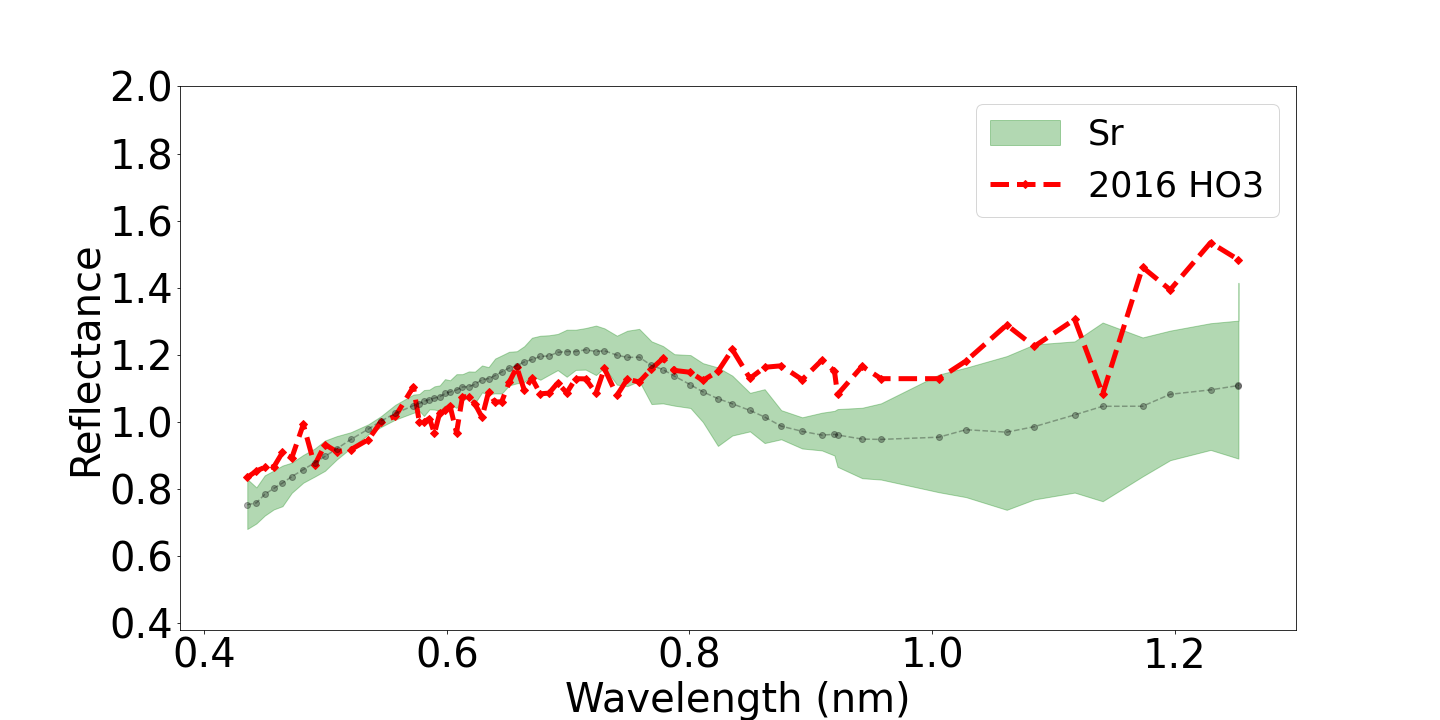}
\includegraphics[width=0.22\columnwidth]{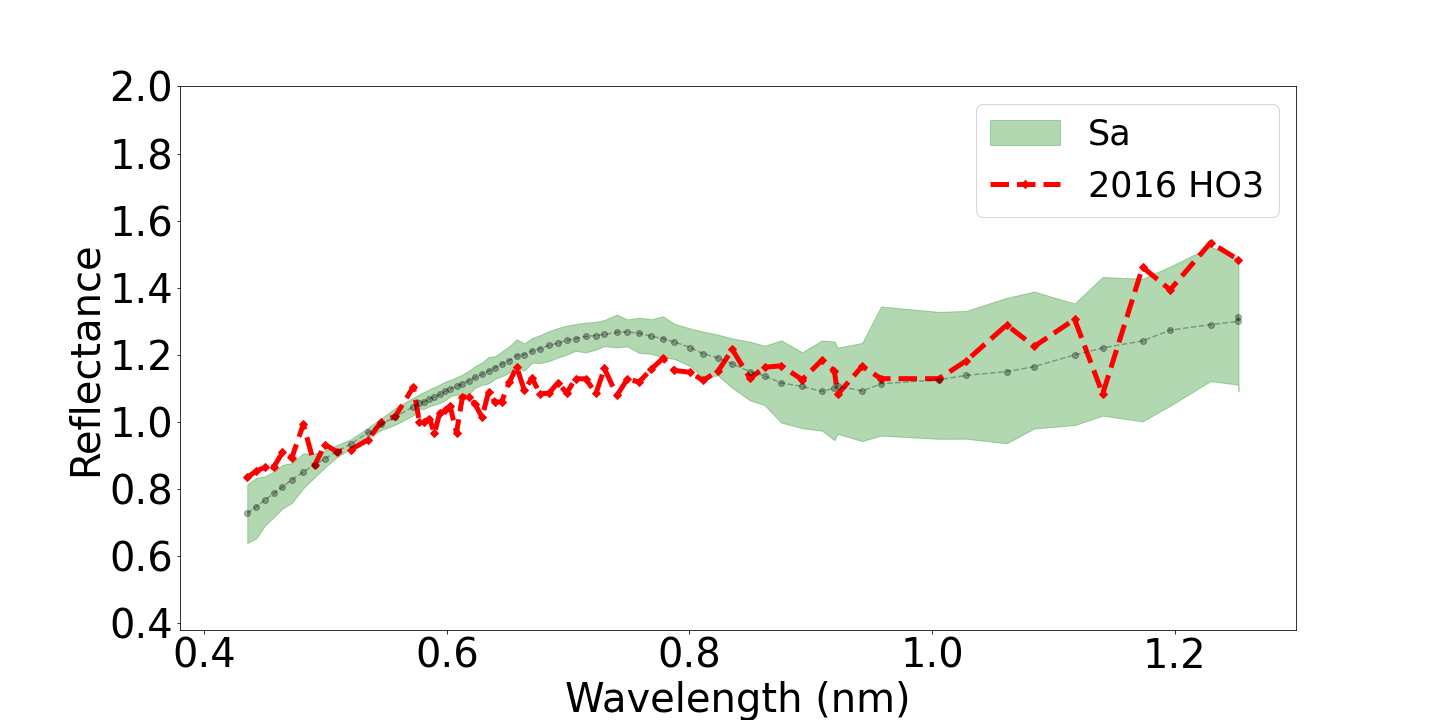}
\caption{Comparing the Kamo'oalewa's spectrum to the training data.}
\label{ANNresult}
\end{center}
\end{figure*}

\subsection{Taxonomy of Kamo'oalewa}
Inputting Kamo'oalewa's spectrum into the built ANNs, the output maximum possibility points to the S-type. Figure~\ref{ANNresult} presents Kamo'oalewa's spectrum (drawn with red color) and the training data (green color) of 8 involved types. It is easy to note that the Kamo'oalewa's spectrum locates almost in the scope of the training data with S-type (the upper right plot of Fig.~\ref{ANNresult}), has a very shallow absorption around the 1-micron band and a redder slope compared to the training data with S-type. The secondary largest possibilities point to the S$_a$ and S$_{qw}$ types. The S$_a$ asteroids are believed to be space-weathered A-type asteroids \citep{DeMeo2009}. Assuming Kamo'oalewa is a fragment of and Atype asteroid, its surface may have an olivine-rich mineralogical composition. Comparing the spectra of Q, S$_q$, S$_{qw}$ and S asteroids, a enhancing trend of space weathering is shown (see the upper row of Fig.~\ref{ANNresult}). Without limiting to the S-type, we also consider that Kamo'oalewa can be a fragment of Q asteroids. Providing it comes from a Q-type asteroid, the time scale of space weathering would be thousands of years, longer than the estimated space-weathering age of Itokawa \citep{Hasegawa2019}. Refering to the composition of Q-type near-Earth asteroids,  Kamo'oalewa could be composed of olivine mineral (40-60 percentage of mass), pyroxene mineral (20-30 percentage), and some iron-nickel metal. Besides the 1-$\mu$m absorption feature in the spectrum, there seem to be shallow absorption around 0.7~$\mu$m, which implies hydrated minerals on Kamo'oalewa's surface. If this is true, the parent asteroid of the Kamo'oalewa may come from the inner solar system.

\section{Thermal inertia of Kamo'oalewa} \label{subsec:thermal inertia}

After the photometric inversion, especially having Kamo'oalewa's spin parameters, shape and absolute magnitude, we re-estimated its thermal inertia based on our new derived $A_2=-13.29349563\times10^{-14}$au/d$^2$ with the signal to noise ratio of $10.32$. This new derived $A_2$ gives an orbital drift of $da/dt=(-56.914\pm5.0)\times10^{-4}$ auMy$^{-1}$. To estimate the thermal inertia of Kamo'oalewa, we applied the ASTERIA (Asteroid Thermal Inertia Analyzer) tool \citep{Fenucci2023,Novakovic2024}. The ASTERIA code applies a Monte Carlo approach to estimate the thermal conductivity $K$ with the linear theory of the Yarkovsky effect \citep{Vokrouhlicky1999} based on the measured Yarkovsky drift $(da/dt)_m$. The thermal inertia $\Gamma$ was actually computed from $K$ by the relation $\Gamma=\sqrt{\rho KC}$.

The model of the Yarkovsky drift is on the left of Eq.~\ref{orbital-drift} (below), combined with the seasonal and diurnal effects, and the measured Yarkovsky drift is on the right:
\begin{equation} 
 \begin{array}{l}
    \frac{da}{dt}(a,e,D,\rho,K,C,\gamma,P,\alpha,\epsilon)
    =(\frac{da}{dt})_m.\\
\label{orbital-drift}
 \end{array}
\end{equation}
The parameters involved ($a, e, D, \rho, K, C, \gamma, P, \alpha, \epsilon $) are the orbital semi-major axis and eccentricity, diameter, bulk density, thermal conductivity, heat capacity, obliquity of spin pole, spin period, absorption coefficient, and the emissivity. The diameter $D$ of Kamo'oalewa was derived from the relation $D=\frac{1329}{\sqrt{p_v}}10^{-0.2H}$, using the derived absolute magnitude $H=24.98$~mag and the assumed geometric albedo $p_v=0.24$. With the new derived pole solution, we computed the obliquity as $\gamma=113$ degree. The prior probability density distributions for parameters $H, p_v, \gamma, P$ and $da/dt$ were generated by using the derived values and their uncertainties. The parameters $\rho, C, \alpha, \epsilon$ were referred to that in the literature \citep{Fenucci2025}. Considering the elongated shape of Kamo'oalewa, we applied a correction coefficient of $\xi=0.75$ for modeling the Yarkovsky orbital drift. Detailed information is listed in Table \ref{pdf_ini}. With an MC procedure, posterior distributions of involved parameters  were derived by scanning the $K$ with a given step and randomly sampling other parameters from their prior probability density distributions.

\begin{table}
\caption{Values of parameters of the Kamo'oalewa used to generate a prior PDF in the thermal inertia estimation.}
\label{tab:thermal_parameter}
\centering  
\begin{tabular}{ll }    
\hline\hline       
Parameter & Value  \\ 
\hline
Absolute magnitude, H (Mag) & 24.98 $\pm$ 0.76 \\
Albedo, $p_V$ & 0.24 $\pm$ 0.05 \\
Bulk density, $\rho$ (Kg $m^{-3}$) & 2720 $\pm$ 540 \\
Rotation Period, P (hour) & 0.474195 $\pm$ 0.0001  \\
Obliquity, $\gamma$ & 113.56 $\pm 5^{\circ}$  \\
Orbital drift, $\frac{da}{dt}|_{obs}(My^{-1}$ ) & -0.0056914 $\pm$ 0.00047\\
Heat capacity, C (J k$g^{-1}$ $K^{-1}$)& 800 \\
Absorption coefficient,$\alpha$ & 0.95  \\
Emissivity, $\epsilon$ & 0.984 \\
Non-sphericity correction, $\xi$  & 0.75 \\
\hline
\label{pdf_ini}
\end{tabular}
\end{table}

The derived posterior distributions of the diameter $D$ and the density $\rho$ show  unique peaks, while the thermal conductivity $K$ and thermal inertia $\Gamma$ present two peaks. To get the best values and uncertainties for the involved parameters, we did the peak fitting to the parameter distributions with the Gaussian function or the Pseudo-Voigt function. The best values of the diameter and density of Kamo'oalewa are 27.88~m and 2692~Kg~m$^{-3}$, respectively. If taking the first peak value of $K=0.0012$, the computed thermal inertia is 48~J~m$^{-2}$ K$^{-1}$s$^{-1/2}$, which is a typical value for main-belt asteroids with surface covered by fine regolith. Here, we prefer the secondary peak of $K=0.0076$, which gives the value of thermal inertia $\Gamma$ around 163.0~J~m$^{-2} $K$^{-1}$s$^{-1/2}$. The best values and uncertainties of the four parameters $K, \Gamma, \rho,$ and $D$ are listed in Table~\ref{thermal_result}. Considering Kamo'oalewa as an S-type near-Earth asteroid, we prefer the thermal inertia value of $163.0^{-33.9}_{+99.2}$ Jm$^{-2}$K$^{-1}$s$^{-1/2}$. Comparing our derived thermal inertia to that of \cite{Fenucci2025}, our result is very close to the result of model 4.  Such a value of $\Gamma=163.0$~Jm$^{-2}$ K$^{-1}$s$^{-1/2}$ means the surface is a mixture of sand grains and crushed boulders. 

\begin{table}
\caption{Estimated Parameters of the Kamo'oalewa with the ASTERIA tool.}
\label{tab:thermal_result}
\centering  
\begin{tabular}{lll}    
\hline\hline       
Parameter & $p_v=0.24$ \\ 
\hline
Conductivity, $K$ (W $m^{-1}k^{-1}$) & $0.0078^{-0.0017}_{+0.0084} $\\
Thermal inertia, $\Gamma$ (J $m^{-2} K^{-1} s^{-1/2}$) & $163.0^{-33.9}_{+99.2} $\\
Bulk density, $\rho$ (Kg$m^{-3}$) & $2692.0^{-208}_{+192}$\\
Diameter, D(m) & $27.88^{-4.2}_{+4.6}$\\
\hline
\label{thermal_result}
\end{tabular}
\end{table}

\begin{figure}[h!]
\begin{center}
\includegraphics[width=0.45\columnwidth]{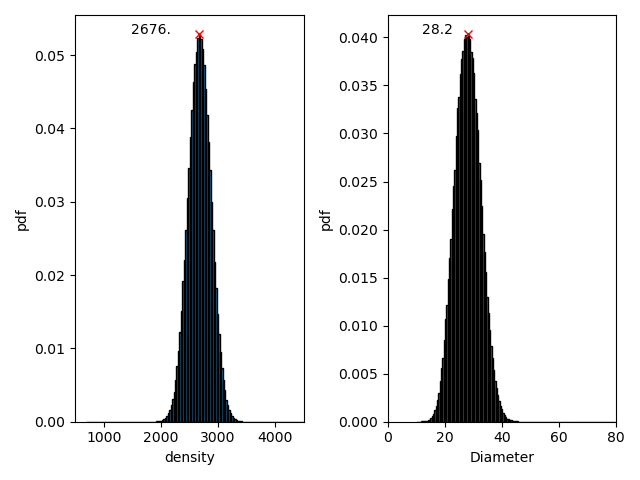}
\includegraphics[width=0.45\columnwidth]{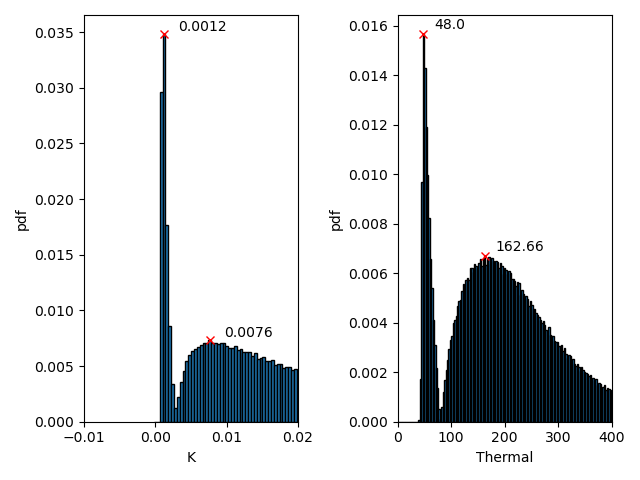}
\caption{The distributions of parameters density, diameter, thermal conductivity and thermal inertia.}
\label{ASTERIA-result}
\end{center}
\end{figure}

\section{Summary} \label{sec:sum}
In all, 17 nights of photometric data of Kamo'oalewa obtained between 2016 and 2018 are used for the photometric inversion with two methods (M-inversion and K-inversion). In the  M-inversion procedure, the relative photometric data set and a combined data set (17 relative lightcurves plus 29 absolute photometric data points) are analyzed by a least-squares algorithm and an MCMC approach, respectively. The least-squares solutions of the spin pole for two datasets are $(276^o.79, -21^o.43)$ and $(277^o.12, -16^o.31)$ with a very close period of 28.45170 minutes. The convex shape from the relative data set is more flat than that from the combined data set. The pole of Kamo'oalewa given by the MCMC procedure is $(287^{o}.02^{-6.70}_{+5.97}, -31^{o}.89^{-2.27}_{+3.04})$, its spin period is $28.4517\pm2.0E^{-7}$ minutes. Based on the posterior distributions of the absolute magnitude and photometry slope of Kamo'oalewa, we estimate its absolute magnitude and photometry slope as $24.98 ^{-0.17}_{+0.18}$ mag and $0.94^{-0.24}_{+0.18}$ mag/rad. Assuming a geometric albedo of 0.24, its diameter is $\sim$30~m. 

As comparison, a pole of $(273^o.58, -20^o.26)$ with a spin period of 28.4517 minutes is derived with the K-inversion using the same relative data set, which is close to values of the M-inversion in the ranges of pole uncertainty.  Its corresponding convex shape (ellipsoid fit: $1:0.80:0.52$) is more close to that from the relative data set in the  M-inversion, except more flat.  Such differences among the convex shapes of three cases, may be risen from the degeneration of the pole latitude and length of C-axis,or the different scattering laws used.  

Based on the available spectroscopic data of Kamo'oalewa, we analyze its taxonomy with the constructed ANN tool, in which 8 types (Q, R, V, S, S$_q$, S$_{qw}$, S$_r$, and S$_a$) are involved. The ANN result suggests a S-type composition of the Kamo'oalewa. Also, we think it may be one fragment of A-type or Q-type asteroids, considering its part spectral data is in the range of the S$_a$, S$_{qw}$ and S$_q$-type's training data, and the fact that  the Kamo'oalewa has undergone the stronger space weathering as a near Earth asteroid. Additionally, a weak absorption around 0.70~$\mu$m implies the existence of hydrated minerals over Kamo'oalewa surface. We guess the parent asteroid of Kamo'oalewa could come from the inner part of the solar system.

The re-estimated thermal inertia of 163.0~Jm$^{-2}$K$^{-1}$s$^{-1/2}$ based on new derived physical parameters and Yarkovsky shift $da/dt$, reflects the Kamo'oalewa has a mixed surface of grains with small boulders, like the surface of asteroid Bennu.

\section{Acknowledgements}

This work is supported by the National Natural Science Foundation of China under grant No. 12288102. We also thank the financial support from the National Natural Science
Foundation of China under grant No. 12373069 and the Research Council of Finland (grants No. 345115 and 336546). We acknowledge the science research grants from the China Manned
Space Program with No. CMS-CSST-2025-A16 and No. CMS-CSST-2025-A20, the Foreign Experts Project (FEP) of State Administration of Foreign Experts Affairs of China (SAFEA) with No. H20240864, and 2025 Annual project for the Institute-level Cooperation program between the Chinese Academy of Sciences and Research Council of Finland. KM's research supported by the Research Council of Finland grants 366954, 359893, and 336546.

\bibliographystyle{aasjournalv7}
\bibliography{HO3}

\end{document}